\documentclass[aps,prb,floatfix,superscriptaddress,twocolumn,amsmath,amssymb]{revtex4-2}

\usepackage{amssymb}
\usepackage{amsmath}
\usepackage{graphicx}
\usepackage{epsfig}
\usepackage{appendix}
\usepackage{bbold}
\usepackage{dsfont}{\tiny }

\usepackage{color}

\def\bea{\begin{eqnarray}}
\def\eea{\end{eqnarray}}
\def\nn{\nonumber}
\def\beq{\begin{equation}}
\def\eeq{\end{equation}}
\def\ba{\begin{eqnarray}}
\def\ea{\end{eqnarray}}
\def\be{\ba\displaystyle}
\def\ee{\ea}

\def\ve{\varepsilon}

\def\L{{\mathcal L}}
\def\M{{\mathcal M}}

\newcommand{\la}{\langle}
\newcommand{\ra}{\rangle}

\usepackage{algorithm}
\usepackage[noend]{algpseudocode}

\begin{document}
\title{Constructing k-local parent Lindbladians for matrix product density operators}

\author{Dmytro Bondarenko}
\email{dimbond@live.com}
\affiliation{Stewart Blusson Quantum Matter Institute, University of British Columbia, 2355 East Mall, V6T 1Z4 Vancouver, BC, Canada}
\affiliation{Institut f\"ur Theoretische Physik, Leibniz Universit\"at Hannover, Appelstr. 2, DE-30167 Hannover, Germany}

\newcommand{\pol}[1]{\textcolor{blue}{#1}}
\newcommand{\del}[1]{\textcolor{red}{#1}}

\begin{abstract}
	Matrix product density operators (MPDOs) are an important class of states with interesting properties. Consequently, it is important to understand how to prepare these states experimentally. One possible way to do this is to design an open system that evolves only towards desired states. A Markovian evolution of a quantum mechanical system can be generally described by a Lindbladian. In this work we develop an algorithm that for a given (small) linear subspace of MPDOs determines if this subspace can be the stable space for some frustration free Lindbladian consisting of only local terms and, if so, outputs a suitable Lindbladian.
\end{abstract}

\maketitle
\date{\today}
\textit{Introduction.---}Engineering evolution that leads to interesting highly entangled states is a core task of quantum simulation and computation. Arguably the simplest preparation scheme involves designing local evolution to produce steady states. If a system is Markovian~\cite{LindTheo0}, such evolution is described by a Lindbladian~\cite{LindTheo,LindTheo1,LindTheo2}. It can be shown that this is the case for systems that are weakly coupled to a large environment~\cite{QEngiBook, RedfieldLindblad}.

Dissipation is often treated as an adversary in this endeavour. However, it can also be a resource. Preparation by dissipation is very robust---unlike in some closed systems, the produced state can be close to the desired one even if the evolution operator has substantial errors; see~\cite[Theorem 4]{CPTP_ergodic} for the case when the stable space is one-dimensional, and \cite[Theorem 7]{Dissipate_stability} for the case of polynomially decaying errors in rapidly-mixing Lindbladians, and compare with orthogonality catastrophe for non-dissipative Fermi gases~\cite{Orthogonality_catastrophe,Orthogonality_catastrophe_generalized}. Local dissipation and driving can also rectify currents~\cite{Dissipation_current_rectification}, activate approximate conservation laws~\cite{Dissipative_currents_integrable} and improve sensors~\cite{Meghana_thesis, Dissipative_sensing}. Stability and simplicity makes this protocol a conducive candidate for quantum memories~\cite{QMemory_dissipation}. There are large classes of states that can be prepared dissipatively, such as stabilizer spaces~\cite{LindEngi1} and, importantly, any state with a so-called parent Hamiltonian~\cite{Dissipation_parentH}.

A Hamiltonian $H$ is called parent for a certain space of states $GS$ if $GS$ is its ground space and $H$ is local, frustration free and gapped. One application of such a construction is to obtain Hamiltonians for novel integrable models. This was one of the original motivations to introduce tensor networks to approximate quantum states. The construction of a Hamiltonian from what is now considered a prototypical matrix product state (MPS) occurred in~\cite{AKLT1} for the AKLT state~\cite{AKLT0}. This result, together with~\cite{FCS}, has inspired a series of generalisations~\cite{AKLT2,AKLT3} culminating in a general way to produce an integrable model from an MPS~\cite{MPSTheo}. The proof that any translational-invariant matrix product state has a parent Hamiltonian was established in~\cite{Wielandt} (see~\cite{ParentHThesis} for a detailed discussion), building on earlier results obtained in~\cite{ParentH2,ParentGap,MPSTheo}. Thus, an MPS can be prepared via local interactions and dissipation~\cite{LindEngi}.

\begin{figure}[t]
	\centering
	\includegraphics[scale = 1]{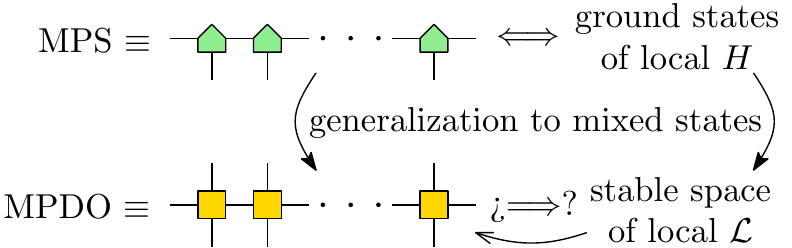}
	\caption{MPS are connected to ground states of local gapped 1d Hamiltonians. MPS can approximate such ground states well, and each MPS is the unique ground state of a parent Hamiltonian. MPDOs and Lindbladians are a direct generalization of MPS and Hamiltonians to mixed states. MPDOs can represent thermal states for high enough temperature and are a well-established variational method for stable states of open-system dynamics. But do MPDOs have parent Lindbladians?}
	\label{fig:L-MPDO} 
\end{figure}

This is a significant finding, as MPSs are used in various algorithms (sometimes quantum, e.g.~\cite{QMPS, QMPS1,QMPS2}), pioneered by the
DMRG~\cite{DMRGwhite,dmrg}. As of today, MPSs are a variational class of choice for the analysis of a large number of one-dimensional systems, see e.g.~\cite{orus2019tensor}; they have also found applications in quantum metrology~\cite{MetroTN}, quantum error correction~\cite{MERACodes,TNCodes,TNdecoding,farrelly2021localTNcodes,cao2021TNcodesLEGO}, category theory~\cite{Categories_intro_our} and machine learning~\cite{TNML,TNML2,TNMLLib}. 
The wide-spread adoption is not a coincidence, as it was proven by Hastings~\cite{hastings} (see~\cite{MPSGS} for the proof with exponentially better parameters) that ground states of gapped Hamiltonians in one spatial dimensions can be approximated arbitrarily well by an MPS in an efficient manner.

Matrix product density operators (MPDOs) are a direct generalization of MPSs to mixed states. While somewhat trickier to work with than MPSs~\cite{MPDOTheoNP,DimBondPurification,MPDOpurification,MPDOpurificationTInv}, MPDOs are at the core of various numerical methods~\cite{DimBondPurificationApprox,QTrajectories,QTrajectories0,MPDONumV,MPDONum,QTrajectoriesVsTimeEvo,MPDOpurificationVsDMRG}). These methods have been successfully used for the numerical investigation of various physical systems (see e.g. ~\cite{MPDOApplSchw1,MPDOApplSchw2, MPDOApplMBL, MPDOApplQuench, MPDOApplImp,MPDOApplBH,MPDOApplThesis,MPDONum5,MPDO_photonic_delay}). See e.g.~\cite{MPDOReview} for a review of methods based on MPDOs and a discussion of alternatives. There are open-source software packages that can be used for such simulations, e.g.~\cite{MPDOLib1,MPDOLib2}.

Just like for MPSs, the success of MPDOs stretches beyond numerical evidence. It is established that MPDOs can represent thermal states of local Hamiltonians for sufficiently high temperatures~\cite{MPDOparentT,ThermalPEPO, LocalHamiltonian,MPDONumWhite,MPDOThermal}.

We seek to extend the parent Hamiltonian construction to mixed states. 
As achievable locality is often limited in experiments, we will concentrate on $k$-local Lindbladians.
We present an algorithm that determines whether there exists a parent Lindbladian for a given space of MPDOs and, if so, outputs such a Lindbladian (see Fig.~\ref{fig:L-MPDO}). Unlike~\cite{MPDOparentT}, we require parent Lindbladian to be exact and do not restrict ourselves to thermal states.
This construction can be used to design preparation protocols for useful MPDOs, provide a way to invent new integrable systems and give insight into what kind of systems are well described by this variational class.

After introducing some necessary preliminaries, we outline the two main steps of the algorithm. We proceed with an in-detail discussion of these steps, followed by a set of examples.

\textit{Preliminaries.---}
Powerful methods for highly entangled quantum many-body systems can be build around approximating states and operators by a contraction of tensors built from variational parameters. Such a variational class can be depicted as a diagram where each tensor corresponds to a shape and each index---to a line. A line shared by two shapes depicts a contraction of the index. 

MPSs are a prominent example of physically relevant states that can be written as a chain of rank-$3$ tensors. Consider $n$ spin-$j$ particles on a line. The space of MPSs is spanned by the states of the form
\begin{figure}[H]
	\centering
	\includegraphics[scale = 1]{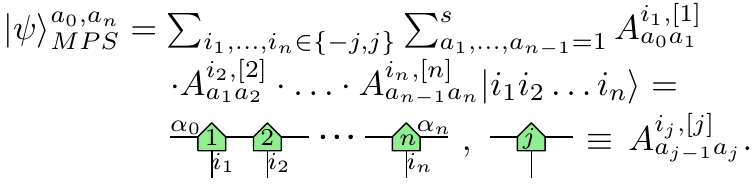}
\end{figure}
In the same spirit as MPSs, MPDOs are defined as density operators that can be written as a chain of rank-$4$ tensors.
\begin{figure}[H]
	\centering
	\includegraphics[scale = 1]{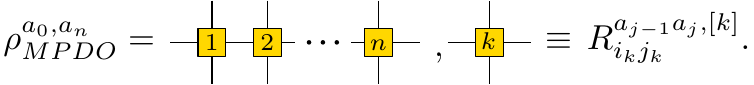}
\end{figure}

Quantum channels are the most general maps physically realizable from density operators to density operators. If the density operator is normalized so that its trace is 1, its eigenvalues $\{p_i\}$ can be interpreted as probabilities of a system to be described by the corresponding eigenvectos $|\psi_i \ra$. This interpretation leads to a requirement, that for any channel $T$, $T\left( \sum_i p_i |\psi_i \ra \la \psi_i | \right) = \sum_i p_i T\left( |\psi_i \ra \la \psi_i | \right)$. Without loss of generality, we can demand that $T$ is linear. As probabilities are positive and add up to one, a channel should be positive---map positive operators to positive operators, and trace preserving. Moreover, one should remember that any system of interest is embedded in the rest of the universe. If we consider a channel that acts as $T$ on a subsystem of interest, and as the identity on some other part of the universe, this map should still be positive. Maps with this property are called completely positive. Quantum channels are defined as completely positive trace preserving linear (CPTP) maps. Any quantum channel can be realized by evolving unitarily some large system and ignoring some of its degrees of freedom, see e.g.~\cite{wolf2012quantum, QInfo}. For a broad overview of how channels can be constructed to produce interesting states, see~\cite{DrD}.

A time evolution of a state $\rho(t)$ from time $t_i$ to time $t_f$ can be described by a quantum channel $T_{t_i, t_f}$. An evolution is called Markovian if $T_{t_i, t_m} \circ T_{t_m, t_f}= T_{t_i, t_f}$ for any $t_i<t_m<t_f$. The generator of such an evolution $\L_t \equiv \lim_{\epsilon \rightarrow 0} \frac{T_{(t+\epsilon,t)} - \mathds{1}}{\epsilon}$ is called a Lindbladian. This yields an evolution equation $\dot{\rho} = \L_t \rho$. 

Of special interest for dissipative preparation are states that do not change with time evolution. Such states, forming the \textit{stable space} of an evolution operator, satisfy $\mathcal{L} \rho = 0$. The stable space is intimately related to the \textit{fixed space} of the corresponding channel. These states naturally arise in long-term dynamics, see e.g.~\cite[Chapter~6]{wolf2012quantum}.

For a quantum channel $T$, the space of states such that
$ \ve(\rho) = \rho $
is called a fixed space. In this text $\mathcal{M}_d$ denotes an algebra of $(d\times d)$-dimensional complex matrices. It is known, see e.g.~\cite[Sec. 6.4]{wolf2012quantum}, that for a trace-preserving positive map
$T:\mathcal{M}_d \rightarrow \mathcal{M}_d$, the fixed space $\mathcal{F}_T \equiv \{\rho| T(\rho)=\rho\}$ is of the form
\begin{equation} \label{tm:Fixed}
\mathcal{F}_T = U \left( \mathbb{0}_{d_0} \oplus \bigoplus_{k=1}^{K} \left( \rho_k \otimes \mathcal{M}_{d_k} \right) \right) U^{\dagger},
\end{equation}
where $U$ is a unitary in $\M_d$, $\mathbb{0}_{d_0}$ is a $d_0 \times d_0$ zero matrix, and $\rho_k$ are diagonal and positive; \mbox{$d = d_0 + \sum_{k=1}^K \mathrm{dim}(\rho_k) \cdot d_k$}.

A channel can also be written in the Heisenberg picture, acting on observables rather than states. Thus, a similar statement to [Eq.~\eqref{tm:Fixed}] exists for algebras of observables. An algebra of observables $\mathcal{I}$ is a linear subspace containing an identity closed under multiplication and Hermitian conjugation. Using the same notation as in [Eq.~\ref{tm:Fixed}], any such algebra can be written as
\be\label{tm:Obs}
        \mathcal{I} = U  \bigoplus_{j=1}^{J} \left( \mathds{1}_{k_j} \otimes \mathcal{M}_{d_j} \right)   U^{\dagger}
\ee
\noindent See e.g.~\cite[Sec. and Theorem 2.7]{CPmapBR} and~\cite[Sec. 1.6]{wolf2012quantum}) for the proof and discussions.

\textit{The algorithm---}There are two important steps in constructing $k$-local parent Lindbladians:
\begin{algorithm}[H]
	\caption{Outline of the algorithm}\label{outlineLMPDO}
	\begin{algorithmic}[1]
	\State For the Lindbladian 
	$\mathcal{L} = \sum_{i=1}^{L-k+1} \mathcal{L}_{i,\ldots,i+k-1} \equiv \sum_{i}\mathcal{L}^{i}$, construct a local term $\mathcal{L}^i$ with stable space \raisebox{-.25\height}{\includegraphics[scale=1]{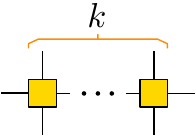}}.
	\State Check that $\mathcal{L} = \sum_{i}\mathcal{L}^{i}$ does not have any additional stable states:
	\begin{figure}[H]\label{JustPatching}
		\centering
		\includegraphics[scale = 1]{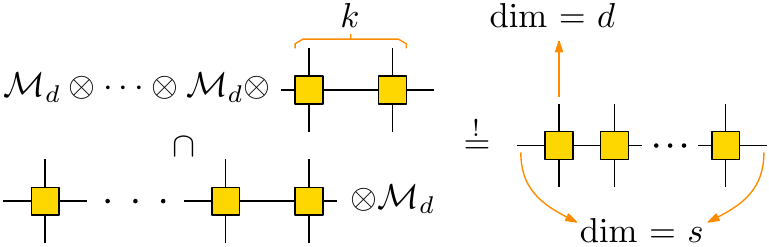}
	\end{figure}
	\end{algorithmic}
\end{algorithm}

Any of these two steps may return a negative result, i.e. either $\mathcal{L}^i$ or $\mathcal{L}$ may have extra stable states in addition to the desired stable state space of MPDOs. In this case, there is no frustration free $k$-local parent Lindbladian for the given space of MPDOs.

\textit{Constructing local term in a Lindbladian.---} 
In this section we will construct a local term in a Lindbladian such that a given length-$k$ space of matrix product operators (MPOs) (see the first step of the Algorithm~\ref{outlineLMPDO})
forms the stable space of the constructed term, or determine that such a local term does not exist. Note that, given a length-$L$ MPDO, the length-$k<L$ MPOs are not necessary density operators. 

To construct a local term with a given stable space $\mathcal{F}$ we will first construct a quantum channel with the fixed space $\mathcal{F}$. For every CPTP map $T$ there exists a Lindbladian $\mathcal{L}$ such that $T[\rho] = \rho \Rightarrow \L[\rho]=0$, for example defined via
\be
		\mathcal{L}[\rho] = \mathrm{const} \cdot (T[\rho] - \rho ) \ 
		\text{for some} \ \mathrm{const}>0.
\ee
Moreover, for the $k$-local and frustration free Lindbladian $\L$, namely
\be
	\mathcal{L} = \sum_{i} \mathcal{L}^i
	\ \text{ and } 
	\left(\mathcal{L}[\rho]=0\quad	\Leftrightarrow \quad \forall i \  \mathcal{L}^i[\rho]=0\right),\nn
\ee
it is possible to associate a $k$-local CPTP map $T$ via
\be
    T_{i,\ldots,i+k-1} &=& e^{\mathcal{L}_{i,\ldots,i+k-1}} \\
	T &=& \frac{1}{L-k+1}\sum_{i=1}^{L-k+1} T_{i,\ldots,i+k-1}
\ee
such that $\L[\rho] = 0 \Rightarrow T[\rho] = \rho$.
This means that 
the construction of a parent Lindbladian and the question of its existence can be investigated by studying fixed spaces of quantum channels.

Let the desired space of $k$-local MPOs starting from site $i$ be spanned by matrices $\left\{O^i_{b}\right\}$, where different values of $b= \{l, r \}$ correspond to different boundary conditions:
\begin{figure}[H]
	\centering
	\includegraphics[scale = 1]{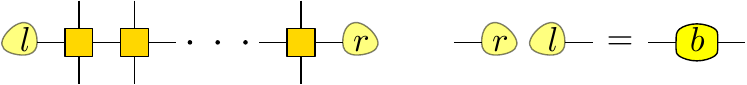}
\end{figure}
Our goal is to find the smallest fixed space $\mathcal{F}$ such that $\forall b \; O^i_{b} \in \mathcal{F}$. If $\dim(\mathcal{F}) = \dim\left(\text{span}\left[ \left\{ O^i_b \right\} \right]\right)$, the local CPTP term $T_{i, \dots, i+k-1}$ can be chosen as a projection on $\mathcal{F}$; otherwise a $k$-local parent Lindbladian does not exist, as for each local term the stable space is too large. Note that any fixed space can be spanned by self-adjoint operators. Thus, it is sufficient to work with self-adjoint linear combinations of MPOs, e.g. 
$\left\{O_j + O_j^\dagger, i O_j - i O_j^\dagger \right\}$.

The algorithm presented below determines the smallest fixed space $\mathcal{F}$ of any CPTP map such that a given set of operators are in $\mathcal{F}$. This algorithm uses subroutines discussed in Appendix~\ref{subsec:BlockDiag}.
\begin{algorithm}[H]
	\caption{Pseudocode for the algorithm that finds the smallest fixed space of a quantum channel that contains a given set of states.}\label{alg:FixedSpace}
	
	\begin{flushleft}
		\textbf{Input:}
		A set of matrices $\left\{ O^i_b \right\}$.
		\\
		\textbf{Output:}
		With probability 1, the smallest fixed space $\mathcal{F}^i$ such that $\forall b \; O_b \in \mathcal{F}^i$. With probability 0, an algebra containing $\mathcal{F}$.
	\end{flushleft}
	\begin{algorithmic}[1]
	\State Find self-adjoint linear independent matrices $\left\{S_i\right\}$ such that $\mbox{$\text{span}\left( \left\{\zeta_i\right\} \right) = \text{span}\left( \left\{O_{i}\right\} \right)$}$.
	\State \label{blockdiagstep}Run the joint block diagonalization algorithm from~\cite{Diag} (Algorithm~\ref{alg:MurotaBlockDiag} in Appendix~\ref{subsec:BlockDiag}). Store the output $Q$ and optional outputs $\{d_i\}, \Sigma$.
	\State Let $D_i= \sum_{j=0}^i d_i$. Construct a matrix 
	\be 
	V = \text{diag}\left(\{ 1 \}_{k=1}^{D_0}, \left\{\sqrt{\frac{1}{\sum_{i=D_0+1}^{D_1} \Sigma_{ii}^2}}\right\}_{k=1}^{D_1}, \dots, \right.\\ \left.
	\left\{\sqrt{\frac{1}{\sum_{i=D_j+1}^{D_{j+1}} \Sigma_{jj}^2}}\right\}_{k=1}^{D_j}, \dots \right).
	\ee
	\State Find the smallest observable algebra~\cite{Diag,Diag0} (Algorithm~\ref{alg:Murota} in Appendix~\ref{subsec:BlockDiag}) for $\left\{V Q^{\dagger} \zeta_b Q  \right\}$. As Algorithms~\ref{alg:Murota} and~\ref{alg:MurotaBlockDiag} share first few steps, use the results already computed during step~\ref{blockdiagstep} of this algorithm. Obtain the set $\left\{U,\{d_j\}_{j=0}^{J},\{k_j\}_{j=1}^{J}\right\}$.
	\State Output the set $\left\{U,V,\{d_j\}_{j=0}^{J},\{k_j\}_{j=1}^{J}\right\}$,
	where $\mbox{$\mathcal{F}^i \equiv U V^{-1} \left( 0_{d_0} \oplus \bigoplus_{j=1}^{J} \left( \mathds{1}_{k_j} \otimes \mathcal{M}_{d_j} \right) \right) U^{\dagger}$}$.
	\end{algorithmic}
\end{algorithm}

Let us run this algorithm for the space $\mathcal{F}^i$ of matrices that span the given space of length-$k$ MPO to get $\mathcal{F}_{T_{i,\dots, i+l-1}}$. If the dimension of $\mathcal{F}_{T_{i,\dots, i+l-1}}$ equals to the dimension of $\mathcal{F}^i$, than we can choose $T_{i,\dots, i+l-1}$ as a projector on $\mathcal{F}_{T_{i,\dots, i+l-1}}$. The local term in a Lindbladian can be chosen as $\mathcal{L}_{i,\dots, i+l-1} = \mathds{1} - T_{i,\dots, i+l-1}$. Otherwise, there is no $k$-local Lindbladian such that its stable space is the desired space $\mathcal{F}^i$.

\textit{Patching local parts - dimension of a stable space} 
In this section we will investigate what happens when $k$-local parts get patched together to form the space of arbitrary long MPDOs (see the second step of the Algorithm~\ref{outlineLMPDO}).

Let us investigate the $k$-local Lindbladian
\begin{figure}[H]
	\centering
	\includegraphics[scale = 1]{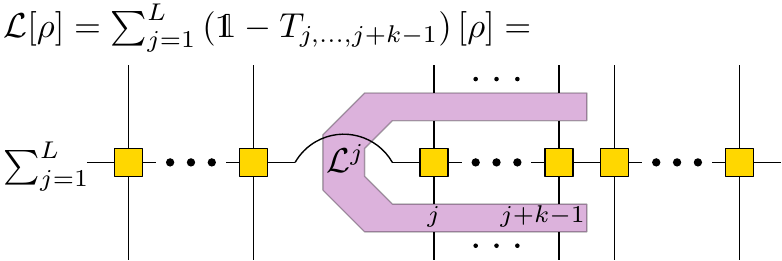}
\end{figure}
\noindent where $T_{j, \ldots, j+k}$ is a CPTP map that acts non-trivially only on $k$ neighbouring qudits. 
By construction, any state that is in the desired stable space of MPDO states is a fixed space of any local CPTP map $T_{j,j+1, \ldots, j+k-1}$. Our goal is to find all states that are in the fixed space of $T_{j,j+1, \ldots, j+k-1}$ for every $j \in \{1,2, \cdots, L \}$.

We will use a technique inspired by \cite{TobiasShore}. Suppose we have a set of $s_n$ linearly independent solutions for the first $l \geq k$ sites in the form
\begin{figure}[H]
	\centering
	\includegraphics[scale = 1]{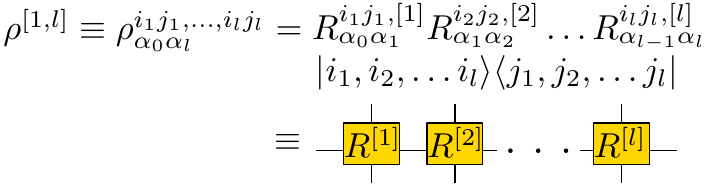}
\end{figure}
\noindent with $i_n, j_n = 1, \ldots d$ and $\alpha_n = 1,...,s_n$; here and below all the repeated indices are summed over. The $R$'s satisfy the linear independence conditions
\be
	R^{i_n j_n, [n]}_{\alpha_{n-1} \alpha_{n}} x_{\alpha_n} = 0, \ \forall i_n, j_n, \alpha_{n-1} \
	\Leftrightarrow \ x_{\alpha_n} = 0, \ \forall \alpha_n
\ee
We now add one more site to the chain and look for eigenvalue zero states of $\mathcal{L}$ in the form
\be
	\rho^{i_1 j_1,\ldots, i_{l} j_l}_{\alpha_0 \alpha_{l}} &=& 
	\rho^{i_1 j_1,\ldots, i_{l-k+1} j_{l-k+1}}_{\alpha_0 \alpha_{l-k+1}} \cdot \nn\\
	&\ \;R&^{i_{l-k+2} j_{l-k+2}, [l-k+2]}_{\alpha_{l-k+1} \alpha_{l-k+2}} 
	\ldots
	R^{i_{l+1} j_{l+1}, [l+1]}_{\alpha_{l+1} \alpha_{l+1}}
\ee

The unknown tensor $R^{i_{l+1} j_{l+1}, [l+1]}_{\alpha_{l} \alpha_{l+1}}$ must satisfy
a system of linear equations
\be
	C_{ \sigma
		, i_{l+1} j_{l+1} \alpha_{l} } 
	R^{i_{l+1} j_{l+1}, [l+1]}_{\alpha_{l} \alpha_{l+1}} = 0, 
\ee
 where $\sigma = \alpha_{l-k+1} i_{l-k+1} j_{l-k+1} \dots i_l j_l$ and
\begin{figure}[H]
	\centering
	\includegraphics[scale = 1]{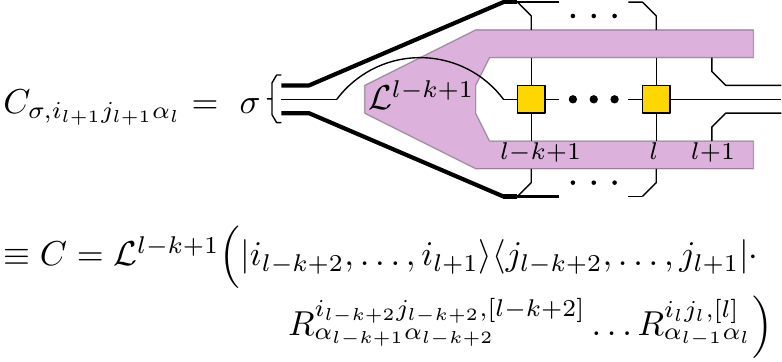}
\end{figure}
\noindent is a matrix with dimension $r s_{l-k} \times d^2 s_l$. The number of solutions is exactly $s_{l+1} = d^2 s_l - \text{rank}(C) $. In particular, this gives a necessary and sufficient condition for the whole stable space to be an space of MPDO form with bond dimension $s$, that is $\text{rank}(C) = s(d^2-1)$.

\textit{Examples.---}
	Let us first illustrate the algorithm on the simplest possible non-trivial sets of states. Consider a set of MPDOs 
	\be
	\rho(\{\alpha_j\}) = \left(\frac{\mathds{1}_2}{2}\right)^{\otimes L} + \sum_{i \in J} \alpha_i \sigma_{i}^{\otimes L},
	\ee
	where $\sigma_i$ are Pauli matrices, $J \subset \{1,2,3\}$ and $\{\alpha_j\}_{j \in J}$ are numbers such that $\rho(\{ \alpha_j \})$ is a state.
	The bond dimension of $\rho(\{\alpha_j\})$ can be chosen as $D = |J| + 1$ with the corresponding tensors
	\begin{figure}[H]
		\centering
		\includegraphics[scale=1]{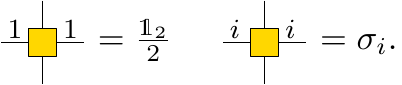}
	\end{figure}
	Let us first look at the case $J = \{3\}$. We will show that 
	$\rho(\{\alpha_3\})$ is the stable space of a two-local Lindbladian.
	Indeed, the first stage of the Algorithm~\ref{outlineLMPDO} outputs a local term $\mathcal{L}_{i,i+1}$ as the projector on the algebra 
	$\mathds{1}_2 \otimes (\mathcal{M}_1 \oplus \mathcal{M}_1)$ minus identity and the second stage proves that the sum of local terms has the same number of stable states as one local term.
	
    However, if $|J| = 2$ there is no $k$-local parent Lindbladian for any $k$. Indeed, the space spanned by $\mathds{1}^{\otimes k},\sigma_{i}^{\otimes k},\sigma_{j}^{\otimes k}$ does not form an algebra for $ i \neq j$, and already the first stage of the Algorithm~\ref{outlineLMPDO} outputs that no desired Lindbladian exists. This situation is rather generic for MPDOs and is drastically different to the case of MPS, where a parent Hamiltonian exists for large enough $k$, see ~\cite{Wielandt,ParentHThesis}.
	
	Beforehand we looked only at sets of density matrices that can be simultaneously diagonalised. If we now look at the case $|J| = 3$, the local term of a Lindbladian will be the projector on the non-commuting algebra $\mathds{1}_2 \otimes \mathcal{M}_2$ minus identity.
	
	Similar examples can be contracted using higher-dimensional representations of $su(2)$. 
	
	We will now look at the thermal states of local Hamiltonians,
	$\rho = \frac{e^{- \beta H }}{\text{tr} \left[e^{- \beta H } \right]}$. Let us start with the Ising model $H = \sum_i \sigma_z^{i} \sigma_z^{i+1}$. This state can be described as a MPDO with bond dimension~2:
	\begin{figure}[H]
		\centering
		\includegraphics[scale=1]{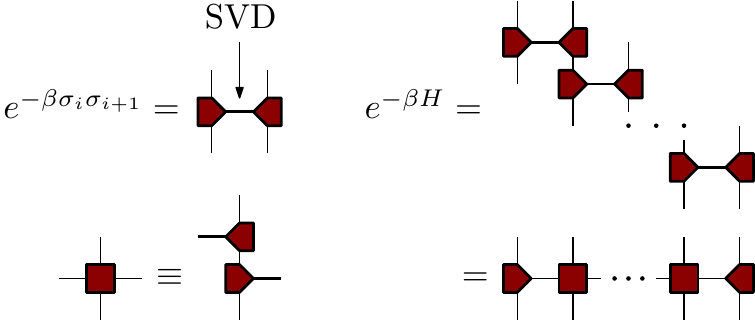}
	\end{figure}
	We are interested in the parent Lindbladian of the space of MPDOs that correspond to the Ising model without the boundary terms
		\raisebox{-.33\height}{\includegraphics[scale=1]{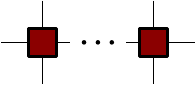}}.
	Let us first try to construct the two-local Lindbladian. The local term can be chosen as the projector on the algebra $\bigoplus_{i=1}^4 \mathcal{M}_1$ minus identity. However, the second stage of the Algorithm~\ref{outlineLMPDO} shows that $\text{rank}(C) = 4 < 6$ and thus, no 2-local parent Lindbladian exists. The situation changes when we increase the locality - the Algorithm~\ref{outlineLMPDO} finds the 4-local parent Lindbladian for the Ising model as a projector on algebra $U\left(\bigoplus_{i=1}^4 (\tilde{\rho}_i\otimes \mathcal{M}_1) \right)U^{\dagger}$ minus identity. Here 
	$\tilde{\rho}_1 = \text{diag}(e^{-8 b}, 1, 1, 1), \ 
	\tilde{\rho}_2 = \text{diag}(1, e^{8 b}, 1, 1), \
	\tilde{\rho}_3 = \text{diag}(1, 1, e^{8 b}, 1), \
	\tilde{\rho}_4 = \text{diag}(1, 1, 1, e^{-8 b})$, $U_{ij} = 1$ if $j=2 i - 1$ or $ i = 8 + j/2$ and 0 otherwise.
	
	While the Ising model plays an important role in physics, it is classical. One way to proceed is too build excitations on top of an entangles state, the other approach is to add non-commuting terms to the Hamiltonian. Let us first  proceed with the later option. The Hamiltonian for the Ising model in a transverse field is $H = \sum_i \sigma_z^{i} \sigma_z^{i+1} + h \sigma_x^i$. The thermal state of this model is entangled, but it might be not so straightforward to write it down in a MPDO form. Thus, we study the first Suzuki-Trotter decomposition of the thermal state. There are two obvious possibilities for such decompositions:
	$\rho_h = \frac{1}{Z_h} \left(e^{- \beta \sum_i \frac{h}{2} \sigma_x^{i} } e^{ - \beta \sum_i \sigma_z^{i} \sigma_z^{i+1}} e^{- \beta \sum_i \frac{h}{2} \sigma_x^{i} }\right)$ and $\rho_\beta = \frac{1}{Z_{\beta}}\left(e^{ - \frac{\beta}{2} \sum_i \sigma_z^{i} \sigma_z^{i+1}} e^{- \beta \sum_i h \sigma_x^{i} } e^{ - \frac{\beta}{2} \sum_i \sigma_z^{i} \sigma_z^{i+1}}\right)$, where $Z_h$ and $Z_{\beta}$ are normalizations. If we denote $\ve = e^{-\beta \frac{h}{2} \sigma_z}$, we can draw the corresponding tensors as
	\begin{figure}[H]
		\centering
		\includegraphics[scale=1]{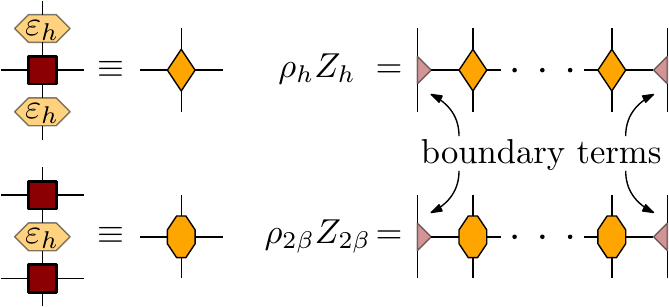}
	\end{figure}
	\noindent We are interested in the space of MPDOs without the boundary therms.
	
	However, the $\rho_h$ is classically correlated, as the bond dimension is $2$.\cite{De_las_Cuevas_2019} For the parameters tested ($\beta =1, h \in \left\{1/10, 1/2, 1, 2, 10\right\}, k \leq 6$) neither of these MPDOs have parent Lindbladians. Every time the first step of the Algorithm~\ref{outlineLMPDO} failed - the stable space of a local Lindbladian is too large.
	
	The other option to generate a quantum model related to Ising model is to start with any wave function 
	$| \psi \rangle$ spanned by $|0\ra^{\otimes L}$ and $|1\ra^{\otimes L}$ and add domain walls at each site with probability $p$ -- for every $n$ starting from $1$ to $L$ we define 
	\be
	\rho_{n-1} = (1-p) \rho_n + p \mathds{1}^{\otimes n} \otimes \sigma_x^{\otimes (L-n)} \rho_n \mathds{1}^{\otimes n} \otimes \sigma_x^{\otimes (L-n)} \nn
	\ee
	starting with $\rho_0 = |\psi\ra \la\psi|$. The space $V_L$ of all $\rho_L$ can be written in MPDO form as
	\begin{figure}[H]
		\centering
		\includegraphics[scale=1]{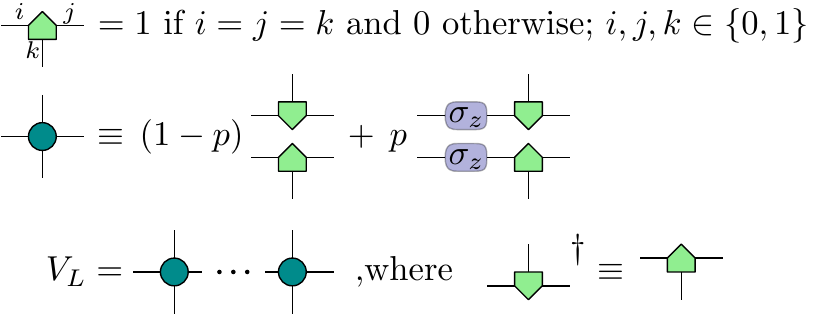}
	\end{figure}
	\noindent 
	The state $\rho_L$ can be entangled (for $p=0$ the $\rho_L$ can be GHZ). Different boundary conditions correspond to different choices of $|\psi\ra$ from a $4$-dimensional space and to mixtures of odd or even number of domain walls. Correspondingly, $\rho_L$ is an element of $8$-dimensional space. There is a 2-local parent Lindbladian MPDOs that give rise to $\rho_L$ and it is the projector on algebra $U(\mathcal{M}_2 \otimes \mathcal{M}_2) U^{\dagger}$ minus identity, where $U$ is a permutation $(2,3,1,4)$ in one-line notation. Note that this algebra is non-commuting.
	
	The code for all of the examples can be found at~\cite{Code_L-MPDO}.
	
\textit{Conclusions and outlook.---}
We developed an algorithm that determines if a given (small) linear space of MPDOs can be a stationary space for some $k$-local Lindbladian and, if so, outputs such a Lindbladian. Such Linbladian exist if and only if a local frustration-free quantum channel with given MPDOs as the fixed space exists.
This gives a recipe for experimental preparation of MPDOs via dissipative engineering.
It is also a good starting point to invent new integrable systems.

The most computationally expensive step of the presented algorithm is simultaneous block-diagonalization of $s^2$ matrices of dimension $d^k \times d^k$, where $s$ is the bond dimension and $d$ is the dimensionality of the underlying elementary physical systems. It scales quadratically with the bond dimension, but exponentially with the locality of the desired Lindbladian.

The interesting remaining questions are what happens to the gap of such Lindbladian in thermodynamic limit and what is a sufficient locality parameter $k$ such that a parent Lindbladian exists. Moreover, it would be useful to understand what kind of MPDOs have parent Lindbladian; it was conjectured by Sofyan Iblisdir that this question might be connected to ranks of different decompositions~\cite{DimBondPurification,DimBondPurificationApprox}. It would be also interesting to study how the ideas from this work can be applied to the random MPDOs and random Lindbladians.

\textit{Acknowledgement.---}The author is thankful to Frank Verstraete, David Perez Garcia, Alexander Nietner, Reinhard Werner, Terry Farrelly, Albert Werner, Gemma De las Cuevas, Kyrylo Snizhko, Sofyan Iblisdir, Luca Tagliacozzo and Polina Feldamann for useful discussions. The author feels especially thankful and indebted to Tobias Osborne for initiating this research project and initial guidance. The author acknowledges support by the SFB 1227 ``DQ-mat'', Project No. A06, of the Deutsche Forschungsgemeinschaft (DFG, German Research Foundation) and Canada First Research Excellence Fund, Quantum Materials and Future Technologies Program.

\bibliographystyle{ieeetr}
\bibliography{references}

\newpage
\begin{appendices}
\onecolumngrid
\section{Simultaneous block diagonalization and subalgebras of observables}\label{subsec:BlockDiag}
From the [Eq.~\ref{tm:Fixed}] in the main text follows, that any fixed space of a quantum channel has a block diagonal structure, $U \left( \mathbb{0}_{d_0} \oplus \bigoplus_{k=1}^{K} \left( \rho_k \otimes \mathcal{M}_{d_k} \right) \right) U^{\dagger} \in U \left( \mathbb{0}_{d_0} \oplus \bigoplus_{j=1}^{\dim(\rho_k)} \bigoplus_{k=1}^{K} \mathcal{M}_{d_k}\right) U^{\dagger}$. The second step of the Algorithm~\ref{outlineLMPDO} in the main text is a subroutine that, given a set of states $\{\rho_i \}_{i\in J}$, finds the smallest fixed space $\mathcal{F}$ such that $\forall i\in J \; \rho_i \in \mathcal{F}$. A reasonable starting point in this endeavor is to simultaneously block diagonalize the set $\{\rho_i \}$, for example via the algorithm presented in~\cite{Diag}.
\begin{algorithm}[H]
	\caption{Pseudocode for the algorithm~\cite{Diag} that finds finest block-diagonalization for a set of self-adjoint matrices.}\label{alg:MurotaBlockDiag}
	
	\begin{flushleft}
		\textbf{Input:}
		A set of self-adjoint matrices $\{ O_i \}$.
		\\
		\textbf{Output:}
		A matrix $Q$ such that $Q^\dagger O_i Q \in \mathbf{Block}\left(\{ d_{i} \}\right)$, where $\mathbf{Block}\left(\{ d_{i} \}\right) = \mathbb{0}_{d_0}\oplus \bigoplus_i \mathcal{M}_{d_j}$. With probability 1, $\dim(\mathbf{Block})$ is minimal.
		\\
		(Optional, for use with in the Algorithm~\ref{alg:FixedSpace} from the main text) The dimensions of blocks $\{d_i \}$,
		a diagonalized random linear combination $\Sigma$ of inputs. 
	\end{flushleft}
	\begin{algorithmic}[1]
		\State Take a random linear combination of the inputs, $A = \sum_i x_i O_i$, where $\{x_i\}$ are uniformly distributed on a real interval, e.g. $[0,1]$.
		\State Find a unitary matrix $R$ that diagonalizes $A$, that is $\mbox{$R^{\dagger} A R=  \Sigma $}$ and $\mbox{$\Sigma  =  \text{diag}\left(\alpha_1 \mathds{1}_{m_1}, \dots \alpha_{k} \mathds{1}_{m_k}
			\right)$}$. Let $R_i$ be a matrix consisting of orthonormal vectors corresponding to eigenvalues $\alpha_i \in \mathbb{R}$.
		\State Put $K= \{1, \dots, k\}$, and let $\sim$ be an equivalence relation on $K$ such that
		\be
		i \sim j \Leftrightarrow \exists p: \; R_i^\dagger O_p R_j \neq \mathbb{0}.
		\ee
		Let $K= \bigcup_i^q K_i$ be the partition of $K$ into equivalence classes with respect to $\sim$. Define matrices $R[K_j] = \left( R_i | i \in K_j \right)$
		\State Output $Q = \left( R[K_1], \dots, R[K_q] \right)$, if needed also $\{d_i \}$ and $\Sigma$.
	\end{algorithmic}
\end{algorithm}
There are several available simultaneous block diagonalization algorithms, see e.g.~\cite{BlockDiag,Diag,Diag0,Diag2,BlockSVD,BlockDiag1,BlockDiagUsePhys,BlockDiag2,BlockDiag3,BlockDiag4,BlockDiag5}. An advantage of the Algorithm~\ref{alg:MurotaBlockDiag} is that it, with probability 1 and in contrast to Jacobi-like algorithms, is exact; moreover, it can be generalised to find the smallest algebras of observables.

Experiments are often composed of a discrete set of machines each measuring a corresponding observable $O_i$. It is useful to understand the power of this measurement apparatus, that is, the smallest algebra that contains the whole set $\{ O_i \}$. This algebra, specified e.g. by a set $\left\{U,\{d_j\}_{j=0}^{J},\{k_j\}_{j=1}^{J}\right\}$ like in [Eq.~\ref{tm:Obs}] in the main text, can be found using the algorithms developed in~\cite{Diag,Diag0}.
\begin{algorithm}[H]
	\caption{Pseudocode for the algorithm~\cite{Diag,Diag0} that finds the smallest observable algebra that contains a given set of observables.}\label{alg:Murota}
	
	\begin{flushleft}
		\textbf{Input:}
		A set of self-adjoint matrices $\{ O_i \}$.
		\\
		\textbf{Output:}
		With probability 1, the smallest observable algebra $\mathcal{I} $ such that $\forall i \; O_i \in \mathcal{I}$. With probability 0, an observable algebra containing $\mathcal{I}$.
	\end{flushleft}
	\begin{algorithmic}[1]
		\State Take a random linear combination of the inputs, $A = \sum_i x_i O_i$, where $\{x_i\}$ are uniformly distributed on a real interval, e.g. $[0,1]$.
		\State Find a unitary matrix $R$ that diagonalizes $A$, that is $\mbox{$R^{\dagger} A R=  \Sigma $}$ and $\mbox{$\Sigma  =  \text{diag}\left(\alpha_1^1 \mathds{1}_{k_1}, \dots \alpha_{d_1}^1 \mathds{1}_{k_1}, \dots,
			\alpha_1^J \mathds{1}_{k_J}, \dots \alpha_{d_J}^J \mathds{1}_{k_J}
			\right)$}$. Let us denote by $R_i^j$ a unitary matrix such that $ \left( R_i^j \right)^\dagger A R_i^j= \alpha_i^j \mathds{1}_{k_j}$.
		\State Let $G_i = (V_i, E_i)$ be a set of directed graphs with vertices $V_i = \{1, \dots, d_i \}$ and edges $E_i = \left\{(l,m;p) : \left(\left( R_l^i \right)^\dagger O_p R_m^i\right) \neq \mathbb{0}\right\}$. Fix a spanning tree $T_i$ for each $G_i$.
		\State For a tree $T_i$, let $P_1^i, \dots P_{d_i}^i$ be the $k_i \times k_i$ matrices that satisfy
		\be
		P_1^i &=& \mathds{1}_{k_i}, \\
		P_m^i &=& \left(\left( R_l^i \right)^\dagger O_p R_m^i\right)^\dagger P_l/c_{plm}, 
		\quad (l,m;p) \in T^i,
		\ee
		where $c_{plm}$ is a positive number such that 
		$\left( R_l^i \right)^\dagger O_p R_m^i  \left(\left( R_l^i \right)^\dagger O_p R_m^i\right)^\dagger = c_{plm}^2 \mathds{1}_{k_i}$.
		\State Output the set $\left\{\tilde{U}^{\dagger},\{d_j\}_{j=0}^{J},\{k_j\}_{j=1}^{J}\right\}$,
		where $\tilde{U}^{\dagger}=R \cdot \text{diag}\left(P_1^1, P_2^1, \dots, P_{d_j}^J \right)$ and
		$\mbox{$\mathcal{I} = \tilde{U}\left[ \mathbb{0}_{d_0}\oplus \bigoplus_{j=1}^{J} \left(\mathcal{M}_{d_j} \otimes \mathds{1}_{k_j} \right) \right] \tilde{U}^{\dagger}$}$.
		\State Perform an appropriate SWAP to get
		$\left\{U,\{d_j\}_{j=0}^{J},\{k_j\}_{j=1}^{J}\right\}$ such that $\mbox{$\mathcal{I} = U \left[ \mathbb{0}_{d_0}\oplus \bigoplus_{j=1}^{J} \left(\mathds{1}_{k_j} \otimes \mathcal{M}_{d_j}\right) \right]  U^{\dagger}$}$.
	\end{algorithmic}
\end{algorithm}
Our Wolfram Mathematica code for this algorithm is available at~\cite{MurotaCode}.
The Algorithm~\ref{alg:FixedSpace} from the main text is the analogue for states and fixed spaces rather than observables.
\end{appendices}

\end{document}